\documentclass[aip,apl,reprint,amsmath,superscriptaddress]{revtex4-1}
\usepackage{graphicx}
\usepackage{amsmath,amsthm,amssymb,latexsym,amsfonts,times,mathrsfs,verbatim, lipsum}
\usepackage{bm}
\usepackage{color}

\begin{document}
\title{Tuning a 3D Microwave Cavity via Superfluid Helium at MilliKelvin Temperatures} 

\author{F. Souris}
\author{H. Christiani}
\affiliation{Department of Physics, University of Alberta, Edmonton, Alberta, Canada T6G 2E9}
\author{J.P. Davis}\email{jdavis@ualberta.ca}
\affiliation{Department of Physics, University of Alberta, Edmonton, Alberta, Canada T6G 2E9}


\begin{abstract}  
Frequency tunability of 3D microwave cavities opens up numerous possibilities for their use in hybrid quantum systems and related technologies.  For many applications it is desirable to tune the resonance at cryogenic temperatures without mechanical actuation.  We show that a superconducting 3D microwave cavity can be tuned at the percent level by taking advantage of the dielectric properties of superfluid $^4$He at milliKelvin temperatures, without affecting its intrinsic quality factor -- reaching $3\times10^5$ in the present experiment.
\end{abstract}

\maketitle

Three dimensional (3D) microwave cavities\cite{Reagor2013} are an important tool in the implementation of quantum technologies.  For example, 3D cavities are one of the most attractive architectures for superconducting quantum circuits, both for their long coherence times\cite{Paik2011} and, more recently, for quantum error correction via cat-states.\cite{Wang2016,Ofek2016}  Furthermore, 3D microwave cavities have played an important role in quantum cavity electromechanics experiments using membranes\cite{Yuan15,Yuan15b,Noguchi2016} and superfluids,\cite{DeLorenzo2014,Singh2016,DeLorenzo2017} as well as quantum cavity magnonics\cite{Goryachev14,Bai15,Lachance2017}  and cavity coupling to solid-state spins.\cite{Probst2014,LeFloch2016} For all of these applications, resonance tunability would be highly beneficial.  Take for example the experiments of Ref.~\citenum{Lachance2017} on coupling a magnonic resonator to a superconducting qubit using a copper 3D cavity.  Since they have tunability of the magnon mode frequency via application of an external magnetic field, they are able to tune through the magnon-qubit avoided level crossing and hybridize the magnon and qubit modes.  But in the case of superconducting 3D cavities, magnetic field tuning is prohibited by the Meissner effect.  Therefore, tunability of the cavity frequency could allow, for example, a qubit to be tuned into resonance with a cavity and observation of Rabi oscillations.\cite{Paik2011}  A second important application of tunable cavities is use as cryogenic microwave filters; phase noise from commercial microwave sources inject unwanted noise into low-frequency cavity optomechanics experiments.\cite{DeLorenzo2014}  Cryogenically-compatible tunable microwave filters would allow higher drive powers, and hence higher optomechanical coupling.\cite{DeLorenzo2014}  Precision resonance tuning is therefore key to advancing 3D microwave cavity technologies.

Often, frequency tuning is performed by mechanically varying the length of a stub inside of a microwave cavity, altering the mode confinement.\cite{LeFloch2013} Yet mechanical actuation is bulky and expensive at the cryogenic temperatures required for superconducting quantum circuits and quantum cavity electromechanics.  Here we present an alternative approach to cavity frequency tuning, by imbibing the cold cavity with a low-loss dielectric, namely superfluid $^4$He.  We show that we can vary the resonance frequency of a superconducting cavity by $2.82$\,\% at milliKelvin temperatures as we fill it with liquid helium, and tune a further $0.26$\,\% by varying the pressure of helium up to $\sim10$ bar.  This approach is fully compatible with quantum circuits and cavity electromechanics, and could have the added benefit of increasing the thermalization of embedded hybrid systems.  

\begin{figure}[b]
\centerline{\includegraphics[width=3.3in]{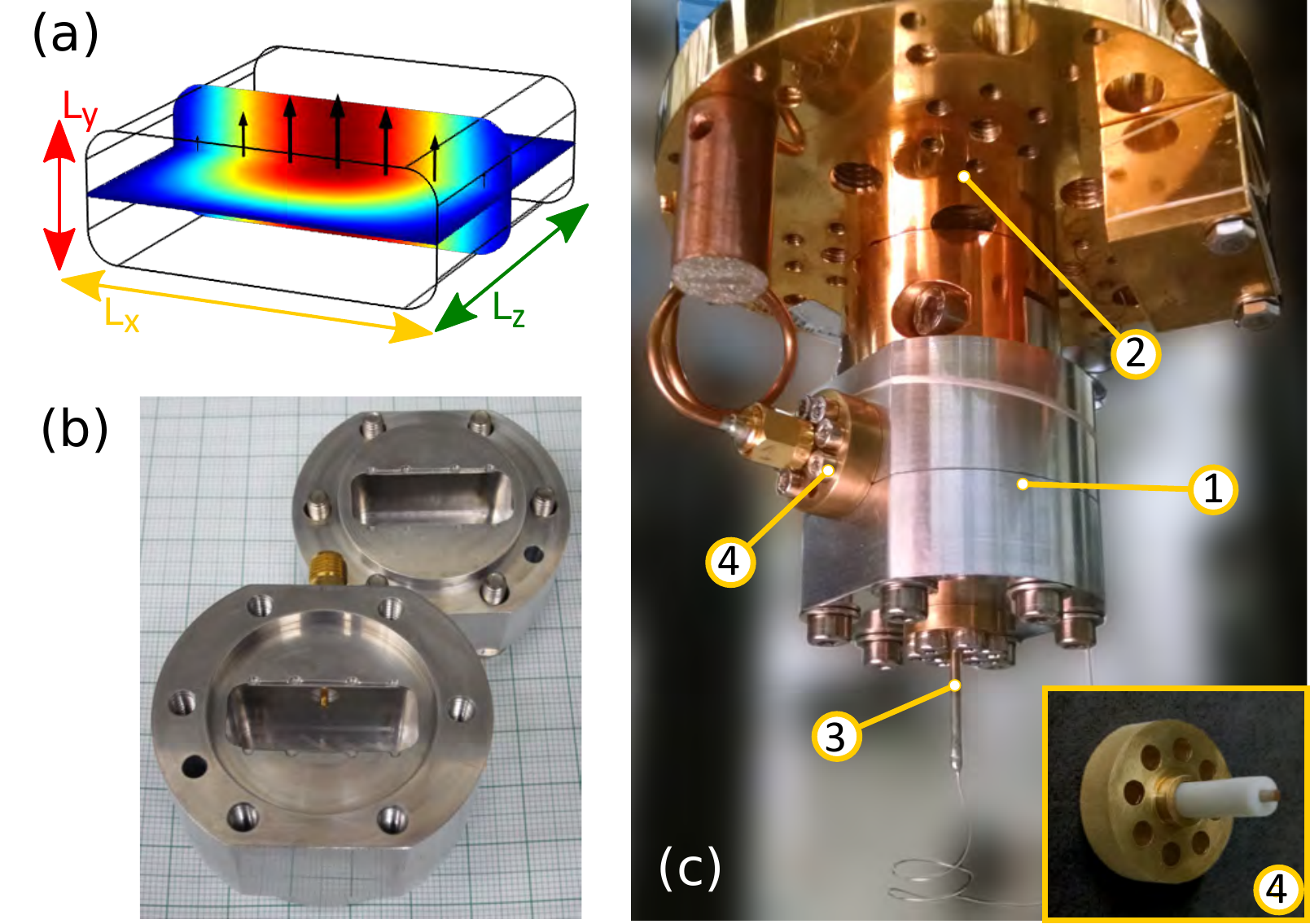}}
\caption{{\label{fig1}}(a) COMSOL simulation of the electric field component of the TE101 mode of our rectangular 3D microwave cavity. The electric field is oriented along the $y$-axis, and is coupled via an extended SMA waveguide.  (b) Photograph of the 3D aluminum cavity, opened to show the internal geometry. (c) Photograph of the assembled 3D cavity (1) mounted to the base plate of the dilution refrigerator (2).  Two brass adapters, which allow a fill-line capillary (3) and a hermetic SMA (4) to be soldered in, are sealed to the cavity using indium o-rings.  The two halves of the aluminum cavity area also sealed using a standard superfluid-leak-tight indium o-ring.  Inset we show an extended waveguide added to the helium side of the hermetic SMA connector, necessary to achieve over-coupling. }
\end{figure}

Our design, Fig.~1, is inspired by the superfluid optomechanics experiments of De Lorenzo and Schwab,\cite{DeLorenzo2014,DeLorenzo2017} although here we use the TE101 mode of a rectangular cavity.\cite{Pozar} This cavity architecture is amenable to integration with qubits\cite{Paik2011} and mechanical resonators via dipole antenna,\cite{Yuan15} or galvanic,\cite{Noguchi2016} coupling.  One convenient aspect of this cavity mode is that it allows for holes and ``splits'' along the lines of surface current, without degrading the intrinsic quality factor.\cite{Cohen17}  Our system includes three such breaks in the cavity body: two for solderable brass adaptors -- one for the fill-line capillary and one for a hermetic SMA connector from Southwest Microwave (Model 290-02G) -- and one break for sealing the two halves of the aluminum cavity.  A second aspect of our cavity is that it is constructed from 6061 aluminum, because of its good performance in superconducting cavities,\cite{Reagor2013} while not being malleable like high purity aluminum or niobium.\cite{DeLorenzo2016}  

The microwave cavity is mounted rigidly to the mixing chamber plate of a dilution refrigerator.  The input drive signal runs from room temperature to the mixing chamber through $48.4$~dB of attenuation.  We measure $S_{11}$ from the cavity using a directional coupler, with $20$~dB of directionality, mounted to the mixing chamber and low attenuation coaxial cables exiting the fridge.  A vector network analyzer is used to probe the microwave cavity, and mode data is fit\cite{Probst2015} to determine the resonance frequency ($f_r$), as well as intrinsic ($Q_i$) and extrinsic ($Q_e$) quality factors. The empty cell resonance frequency at $60$~mK is found to be $f_0=7.9136395$~GHz $\pm$ 2~kHz, close to the $7.85$~GHz calculated for the TE101 mode of a rectangular cavity with dimensions $L_{\textrm{x}}, L_{\textrm{y}}, L_{\textrm{z}} = 27, 10, 27$~mm. We operate in the over-coupled regime, where $Q_e \approx 6\times10^3$ when the cavity is evacuated.  $Q_e$ rises to $8\times10^3$ when filled with liquid helium as a result of the dielectric constant of helium liquid changing the waveguide coupling of the hermetic SMA.  At room temperature the intrinsic quality factor is limited by the conductivity of aluminum 6061.  Below the superconducting transition temperature, $Q_i$ plateaus at an intrinsic quality factor of $3\times10^5$, consistent with literature values.\cite{Noguchi2016,Cohen17}  Unlike $Q_e$, $Q_i$ is entirely unaffected by the presence of liquid helium.

\begin{figure}[b]
\centerline{\includegraphics[width=3.3in]{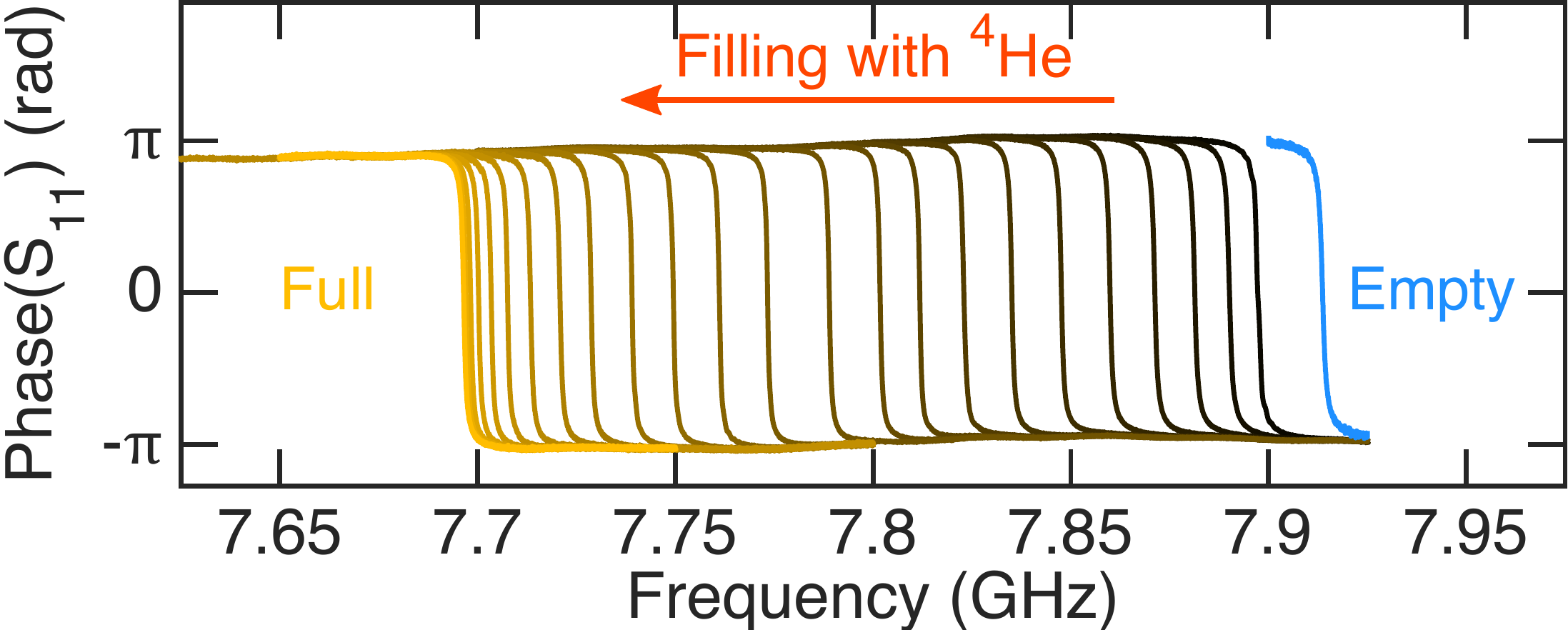}}
\caption{{\label{fig2}} Evolution of the superconducting 3D microwave cavity resonance as it is filled with liquid helium. Each colored curve corresponds to the phase of $S_{11}$ signal measured every $\sim30$~s, the total fill time being about $10$~minutes. While filling, the mixing chamber stays below $100$~mK. Filling could be stopped at any point, thereby setting the cavity resonance frequency.}
\end{figure}

The cavity is filled with helium ($99.995$\,\% chemical purity) \textit{in situ} through a Cu-Ni capillary with $400$~$\mu$m outer diameter and $75$~$\mu$m inner diameter, which passes through custom sintered-copper-powder heat exchangers at every stage of the dilution refrigerator.  A room temperature gas-handling system allows the helium pressure to be monitored and easily adjusted.  Because of the efficient heat-exchangers, the cavity can be filled with helium while the mixing chamber stays below $100$~mK, in as little as 10 minutes.  The filling rate could be controlled by adjusting the helium flow from the gas handling system.  In Fig.~2 we show how the resonance frequency of the TE101 mode evolves as the 3D cavity is filled with liquid helium at saturated vapor pressure.  We see a smooth shift of the cavity frequency from $7.9136$ to $7.6965$~GHz, corresponding to $2.82$\,\%.  This is as expected from the dielectric constant of liquid helium $\varepsilon_\textrm{He}$, since $\sqrt{\varepsilon_\textrm{He}} = 1.0282$ at low temperature and saturated vapor pressure.\cite{Brooks1977}   

While the work of De Lorenzo and Schwab demonstrated the containment of superfluid helium inside a 3D microwave cavity,\cite{DeLorenzo2014} and that the resonance frequency shifted as they filled their cavity at $4$~K,\cite{DeLorenzo2016} they only worked up to saturated vapor pressure.  Here we demonstrate that one can further tune the resonance frequency by pressurizing the superfluid helium \textit{in situ}.  In Fig.~3, we show a shift of up to $0.26$\,\% of the resonance frequency at a constant temperature of 60 mK, as the cavity is pressurized above $10$~bar, with a maximum tunability of $0.53$\,\% set by the melting pressure of solid helium $P_\textrm{m} = 25.32$~bar. The red curve in Fig.~3 is the expected resonance frequency $f_\textrm{He}$ based on calculated values of $\varepsilon_\textrm{He}$, such that $f_\textrm{He} = f_0/\sqrt{\varepsilon_\textrm{He}}$. The dielectric constant $\varepsilon_\textrm{He}$ is calculated using the Clausius-Mossotti relation:
\begin{equation}\label{eq:Eq1}
	\frac{\varepsilon_\textrm{He}-1}{\varepsilon_\textrm{He}+2} = \frac{4\pi\alpha\rho}{3M}.
\end{equation}
The density $\rho$ is extrapolated from tabulated values,\cite{Brooks1977} and the molar polarizability $\alpha$ is derived from the values measured by Kerr and Sherman.\cite{Kerr1970} The deviation between the expected values and those we measure is at the level of ppm.  This minor discrepancy is believed to be a result of slight bowing of the 3D cavity under pressure.  

\begin{figure}[t]
\centerline{\includegraphics[width=3.3in]{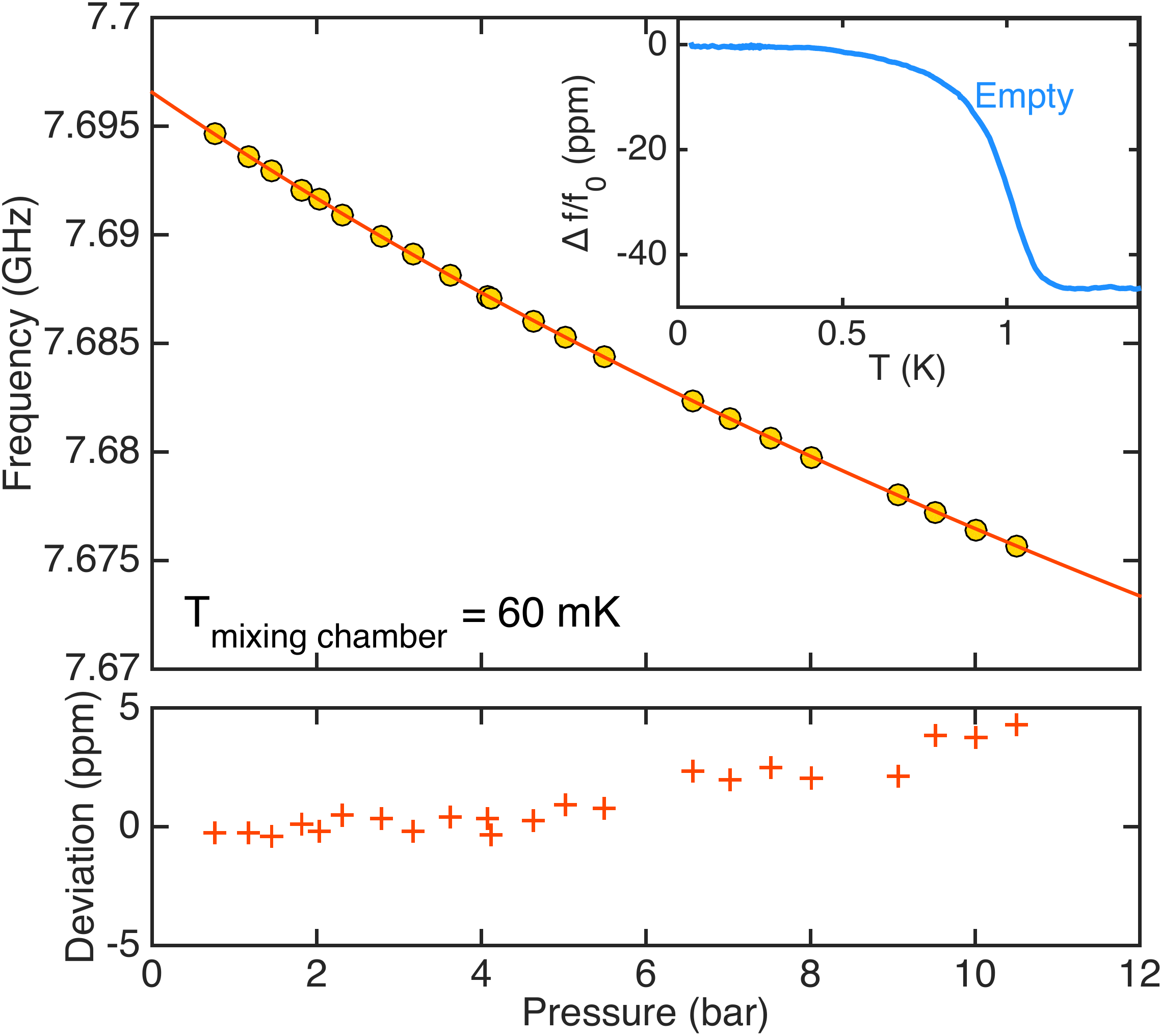}}
\caption{{\label{fig3}} Resonance frequency of the microwave cavity at 60 mK as a function of liquid helium pressure.  The yellow circles are the measured values, and the red line is the expected frequency dependence based only on the initial cavity frequency and the liquid helium dielectric constant calculated from Eq.\,\ref{eq:Eq1}. The deviation between the experiment and the expected values are at the level of ppm.  The deviation grows with increasing helium pressure, likely due to deformations of the cavity geometry from pressurization. The inset shows that below $\sim300$~mK the relative resonance frequency shift $\Delta f / f_0= (f_r-f_0)/f_0$ is below $1$~ppm.}
\end{figure}

The pressure dependence of the cavity frequency, although small, provides a mechanism for frequency stabilization.  For example, we have previously shown the ability to stabilize the pressure of helium to within $2$~mbar using a PID loop.\cite{Rojas2015}  This would correspond to stabilizing the cavity frequency to approximately $4$~kHz.  Significantly better stability would be possible by feeding back from the measured cavity frequency, instead of from an external pressure gauge.  This approach could also be used to stabilize frequency fluctuations from other sources, such as temperature variations -- although these are small below $\sim300$ mK, as seen in the inset to Fig.~3.

In conclusion, we have demonstrated that superfluid helium can be used to tune \textit{in situ} the resonance frequency of a superconducting 3D microwave cavity at milliKelvin temperatures.  By varying the amount of liquid in the cavity, one can tune the resonance frequency by as much as $2.82$\,\%, corresponding to $220$~MHz in the present case.  The frequency can be tuned an additional $0.53$\,\% by pressurizing the liquid inside of the cavity, providing a method of fine tuning the cavity resonance frequency.  This sensitivity to pressure could also be used for stabilization of the cavity frequency.  Importantly, no change is observed in the intrinsic quality factor due to the presence of the liquid helium.  This technique has potential applications in superconducting quantum circuits inside of 3D cavities and quantum cavity electromechanics experiments, where frequency tunability opens up the possibility of exploring hybridization in coupled resonator systems.  In addition, cryogenically compatible frequency tuning has applications in microwave filter cavities to reduce phase noise in quantum cavity electromechanics experiments.  Finally, such superfluid imbibed microwave cavities will enable dipole antenna coupling\cite{Yuan15} to superfluid Helmholtz resonators,\cite{Rojas2015,Souris2017} which opens up the possibility of observing and controlling the motion of a superfluid mechanical mode at the quantum level.

We would like to thank Greg Popowich for his help assembling the dilution refrigerator and for constructing the sintered-copper-powder heat exchangers.
This work was supported by the University of Alberta, Faculty of Science; the Natural Sciences and Engineering Research Council, Canada (Grants Nos. RGPIN-2016-04523 and DAS492947-2016); and the Canada Foundation for Innovation.  H. Christiani thanks the University of Alberta URI for the undergraduate research equipment stipend.

\end{document}